\documentclass[12pt,preprint]{aastex}

\newcommand{\beq}{\begin{equation}}
\newcommand{\eeq}{\end{equation}}

\def\ln{{\rm ln}}

\def\3he{$^3$He\,}
\def\he4{$^4$He\,}

\shorttitle{Elementary Energy Release Events}
\shortauthors{Liu et al.}

\begin{document}

\title{Elementary Energy Release Events in Solar Flares}
%: Spectral Soft-Hard-Soft Evolution}

\author{
%Vah\'{e} Petrosian\altaffilmark{1,2,3}, Yan Wei Jiang
%\altaffilmark{1,2} and 
Siming Liu\altaffilmark{1} 
and Lyndsay Fletcher\altaffilmark{1} 
}
%\altaffiltext{1}{Department of Physics, Stanford University, Stanford,
%CA, 94305 
%email; vahep@stanford.edu; arjiang@stanford.edu}
%\altaffiltext{2}{Kavli Institute for Particle Astrophysics and Cosmology,
%Stanford University, Stanford, CA 94305}
%\altaffiltext{3}{Also Department of Applied Physics}
\altaffiltext{1}{Department of Physics and Astronomy, University of
Glasgow, Glasgow, G12 8QQ, Scotland; sliu@astro.gla.ac.uk}

\begin{abstract}

% Motivated by observations of thermal and nonthermal emissions and
% spectral soft-hard-soft evolution of some hard 
% X-ray pulses of solar flares, we introduce the concept of elementary energy
% release events for phenomenological modelling and discuss the relevant
% physical processes in the context of stochastic particle acceleration by
% magnetised turbulence. 
 Most theoretical investigations of particle acceleration during solar
 flares cannot be applied to observations for detailed study of the
 time evolution.  
%Since charged particles gain energy from the electromagnetic
% fields, the dynamics of the electromagnetic fields is an essential
% element of any self-consistent particle acceleration model.
 We propose a phenomenological model for turbulence evolution
 and stochastic particle acceleration that links observations to the
 energy release and particle acceleration through two coefficients
 characterising particle interactions with turbulent electromagnetic
 fields.  
% Given the large coronal volume involved in
% flares, an energy cascade from large to small spatial scales seems to
% be inevitable, which we attribute to turbulence. 
 In the linear
 regime the particle distribution does not affect the turbulence
 energy cascade. It is shown 
 that electron acceleration critically depends on the intensity of
 small-scale turbulence and an impulsive nonthermal component only
 appears near the peak of the gradually evolving turbulence
 intensity. The model naturally 
 reproduces the 
 soft-hard-soft pattern of hard X-ray pulses, and we attribute the observed
 change in flux and spectral index correlation from the rise to
 decay phase of some pulses to changes in the background
 plasma. Detailed modelling of  
 well-observed individual events will probe the energy
 release processes. 

%We also propose a procedure to identify elementary electron
 %acceleration events from the observed X-ray light curves. A comparison
 %of the elementary energy release and electron acceleration event leads
 %to a well defined and readily measurable electron acceleration
 %efficiency. For flaring corona loops, the model predicts strong
 %correlation between the thermal and nonthermal X-ray emission
 %component, which can be tested with detailed data analyses.  

\end{abstract}

\keywords{Sun: flares --- Sun: particle emission --- acceleration of
particles --- plasmas --- turbulence}

\section{INTRODUCTION}
\label{intro}

% Acceleration rate, especially in terms of power of nonthermal
% particles, site, and efficiency are the essential issues in the 
% solar flare study. The first two are relatively well-constrained by the
% association of observed light curves and images with the
% relevant physical processes (Miller et al. 1997; Aschwanden 2002). 

Given the short particle acceleration timescales inferred from observations of
solar flares, it has been assumed that the acceleration processes are
decoupled from other processes in most theoretical models \citep{m97, a02,
p04, f08}. As a result, theoretical investigations of particle
acceleration are aimed at meeting some 
basic observational requirements, such as timescales, energetics, and
numbers. There are significant ambiguities in quantitative tests of
specific models with observations due to free parameters introduced to
characterise the related, not-well-understood processes. On the other
hand, particle acceleration is just one important aspect of the energy
release processes, which are essentially multi-scale as evident from the
rich time, space, and energy characteristics of flares. Direct
modelling of the energy release, which can be related to
magnetohydrodynamical simulations of large-scale processes, may better %advance
our understanding of flares. In the context of stochastic
particle acceleration (SA) by turbulent electromagnetic fields, we introduce the
concept of elementary energy release events characterised with an
energy release rate and length scale and discuss its implications.

The relation between the gradually varying thermal and bursty
nonthermal emission component is a 
critical aspect of flare studies \citep{v05}. Previous modelling
focused on the correlation between the light-curve of nonthermal
emission and
derivative of the thermal emission light-curve
%. observed in some flares 
\citep{d93}.  
This correlation has been attributed to chromospheric evaporation driven
by a beam of nonthermal electrons that is injected at the top of
flaring coronal loops, propagates to the chromosphere along magnetic
field lines, and collisionally heats the background plasma \citep{b73}. In this
`classical' electron beam model, most of the nonthermal %electron
energy is reprocessed to appear as thermal energy producing the dominant
thermal signatures \citep{n68, a84}. % \citep{a05}. 
 However, this model does not address
the origin of impulsive nonthermal electrons, and has difficulties
in accounting impulsive soft X-ray emission observed from some
footpoints %sources 
and slower than expected decay of thermal emission after
the nonthermal emission already disappears \citep{d93, h94, l97, j06}. By
assuming a low-energy cutoff for a power-law or broken-power-law
nonthermal electron distribution, observations give a very poor  
 constraint on the energetics %of the electron beam 
since 
%the steep power-law
% slope is usually steep so that 
the beam power is dominated by
 low-energy electrons due to steep power-law slopes, and there are
 uncertainties in observationally  
 constraining this low
 cutoff energy due to the dominance of low-energy X-rays by a thermal
 component \citep{h03, s05}. Such a cutoff is also not expected in
 most theoretical models hampering further investigations 
%our understanding of the relevant physical processes 
even for some well-observed flares
 \citep{v05}. 
%Although it is well known that this 
% procedure was introduced for mathematical simplifications \citep{b71},
% the physical implication of a low energy cutoff appears to be
% exaggerated \citep{h03,s05}. The spectral
% flattening of nonthermal hard X-ray (HXR) emission 
% toward low energies is usually considered as evidence for a-low energy
% cutoff of the electron distribution \citep{s07}. Recent detailed
% studies of many flares 
% observed by RHESSI have shown some of this spectral flattening may be
% caused by the photosphere albedo effect \citep{k06, k07}. It
% may also be attributed to a saturation effect of high electron beam flux
% toward the chromospheric footpoints \citep{b77, z06, a07}.  
With the elementary energy release events, we propose a 
%physically motivated 
simple alternative to this electron beam model.
% so that underlying physical processes may be better revealed. 
The model is based on two facts that, 1) while 
 low-energy particles couple strongly with each other through  Coulomb
 collisions forming a thermal distribution, 2) high-energy
 particles must decouple from the thermal background and obtain their
 energy through interactions with electromagnetic fields \citep{h92}. The key
 issues are then the dynamics of the %electromagnetic 
fields and their
 interactions with charged particles.

 A significant amount of energy can be released during flares, and the flux of
 nonthermal particles inferred from observations in combination with the source
 size measurement %from the corresponding images 
sometimes implies a
 nonthermal particle density possibly comparable to that of the coronal
 background plasma \citep{f08}, suggesting very efficient 
 acceleration. 
%So the energy release and particle acceleration
% in solar flares are predominantly macroscopic processes. 
A large fraction
 of the solar corona must be involved if the bulk of the energy and
 accelerated particles are
 stored in the pre-flare coronal magnetic fields, as postulated in most
 theoretical models \citep{g06}. On the other hand, the strong coronal
 magnetic field implies tiny gyro-radii for charged 
 particles.
% in the nearly fully ionised corona 
%even for the very
% energetic particles accelerated during flares. 
%The ion skin depth and the Coulomb collision mean free path are also not long. 
 An energy
 cascade from large to small scales appears to be inevitable.
% so that individual particles can have access to the released energy. 
%There are extensive studies of particle acceleration in a current sheet
 %formed in the free energy release process of magnetic reconnection
 %proposed for flares (Holman 1987; Hannah \& Fletcher 2004; Longcope \&
 %Priest 2007; Wang \& Lapenta 2008). While this process may be able to
 %produce sufficient nonthermal electron fluxes for small flares, a huge
 %number of current sheets would be needed to account for large flares,
 %which essentially invokes an energy cascade from large to small scales
 %implicitly (Benka \& Holman 1992). For models with large scale current
 %sheet, an anomalous resistivity, which must be caused by small-scale
 %turbulence, is usually necessary to generate  sufficiently high
 %electric fields.   
 Turbulence is the most natural agent for energy cascade over a large
 dynamical range, and the corresponding SA
 models have been widely
 used for solar flare studies 
 \citep{m97,p04}. 

 In these models a broad spectrum of particles is energised by
 interacting stochastically with a %prescribed 
spectrum of electromagnetic
 fluctuations over a broad range of spatial and temporal scales.
%However turbulence is a complex highly 
% nonlinear phenomenon, and we still lack a quantitative theory for
% turbulence in magnetised collisionless plasmas, especially that in the solar
% corona. Although magnetised turbulence has been studied continuously
% over the past few decades, especially over the past 15 years, its role
% in the free energy dissipation process remains obscure \citep{m81,
% g95, j09}. Nevertheless, the SA models capture
% the key element of 
One of the key features of the SA is that
wave-particle interactions % rates 
determine not only the energy gain rate
of particles %from the turbulence 
but also their spatial diffusion along
magnetic field lines. Consequently, the particle acceleration is
very sensitive to the turbulence intensity \citep{b77, p04}, which may
explain the bursty behaviour of nonthermal emissions. Detailed studies
of nonthermal hard X-rays (HXRs) also reveal 
%show that the impulsive nonthermal emissions have pulses with 
spectral soft-hard-soft (SHS) evolution of HXR pulses suggesting
independent electron acceleration events \citep{g04, b06}. 
%In this paper, 
We propose a phenomenological model for the evolution 
of turbulence associated with %individual 
elementary energy release events
and use two coefficients to characterise electron
interactions with the turbulence (Section \ref{turbulence}). The model
can naturally fit HXR observations. With 
imaging spectroscopic observations of RHESSI, the model will allow us to
 extract the  
turbulence properties and its evolution for individual
well-observed flares,
which constrain the wave-particle interactions and better our
understanding of the
relation between the thermal and nonthermal emission components (Section
\ref{dis}).  
%for solar flares and parameterise the essential physical processes in
%terms of energy dissipation, plasma heating, particle acceleration and
%transport (\S\ \ref{theory}).  
%The model is applied to solar flares to identify elementary electron
%acceleration events in \S\ \ref{observation}, where we also discuss the
%appropriate procedure to measure the electron acceleration efficiency
%and several predictions on the correlation between the thermal and
%non-thermal emission component. 
Conclusions are drawn in Section \ref{con}. Notations for
different quantities %of the model 
are listed in Table 
\ref{table1}.

\section{Elementary Energy Release Events and Stochastic Particle Acceleration}
\label{turbulence}

\begin{deluxetable}{llrr}
%\tabletypesize{\scriptsize} 
\tablecaption{ Notations   \label{table1}} 
\tablewidth{0pt} \tablehead{ \colhead{Quantities} &
\colhead{Symbol} & \colhead{Typical Value} & \colhead{Units}} 
\startdata
Basic Quantities & & & \\
\hline
Magnetic Field        & $B$         &  100--500              & Gauss     \\
Density      & $n$         &  $10^{9}$--$10^{10}$   & cm$^{-3}$   \\
Flaring Region Length  & $l_0$      &  $10^9$--$10^{10}$     & cm        \\
Loop Cross Section         & $A$    &  $10^{15}$--$10^{16}$ & cm$^2$  \\
Temperature  & $T$       &  $10^6$--$10^7$       & K \\
Energy Release Scale  & $l_e$       &  $10^7$--$10^{9}$     & cm        \\
Energy Release Level  & $b_0$  &  $<1$               &  1         \\
\hline
%Constants & & & \\
%\hline
%Speed of Light        & $c$         &  $3\times10^{10}$ & cm s$^{-1}$ \\
%\hline
Derived Quantities & & & \\
\hline
Alfv\'{e}n Speed      & $v_{\rm A}$ &   $10^8$--$10^9$             & cm s$^{-1}$ \\
Turbulence Intensity      & $b$    &  $\le b_0$        &  1         \\
Eddy Speed            & $v_e$       &  $b$           & $v_{\rm
 A}$     \\
%Electron Energy       & $E$         &  $>1$               & keV       \\
Transition Energy     & $E_t$         &  $ 1$      & $k_{\rm B}T$       \\
Particle Speed     & $v$         &  $>10^{9}$          & cm s$^{-1}$
 \\ 
Wave Transit Time  & $\tau_0\equiv l_e/v_{\rm A}$ &   $0.01$--10 & s \\
Eddy Turnover Time    & $\tau_e \equiv l_e/v_e$ & $b^{-1}$ & $\tau_0$   \\
Rise Time  & $\tau_r\equiv \tau_0/b_0$ &  $b_0^{-1}$ &
 $\tau_0$            \\
Decay Time   & $\tau_d\equiv \tau_e v_{\rm A}/v_e$ & $b^{-2}$ &$\tau_0$ \\
Particle Transit Time & $\tau_t(v)\equiv l_e/v$     & $<0.01$--1    & s$$ \\
%Acceleration Coefficient & $S_a$ & ? & 1 \\
%Scattering Coefficient & $S_s$  & ? & 1 \\
Scattering Time      & $\tau_{sc}(v)$ & $S_sb^{-2}$ & $\tau_t$  \\
Acceleration Time    & $\tau_{ac}(v)$  &$S_ab^{-2} 
%c^2/v_{\rm A}^2
$ &
 $\tau_t$  \\
Escape Time          & $\tau_{esc}(v)$ & $l_0^2v^{-2}\tau_{sc}^{-2}$ &
 $\tau_{sc}$ \\ 
%Coulomb Loss Time    & $\tau_c(E)$ & $0.01(E/{\rm keV})^{3/2}$ & s \\
%Acceleration Efficiency & $\eta_{ne}$ & & \\
\enddata
\end{deluxetable}
Most SA models treat turbulence as an input and do not consider its
 dynamical evolution \citep{b77, m97, p04, g06}. Although \citet{b09}
 modelled the turbulence, their particle acceleration is
 sensitive to an injection process, which is treated as an input
 independent of the turbulence evolution. Observations of solar flares, on the
 other hand, indicate that the large-scale energy release and particle
 acceleration %processes 
are intimately connected. We aim at realising
 such a connection with a phenomenological SA model so that the
 model can be applied to individual events for detailed studies.

  We consider elementary energy
 release events, for which large scale eddies are assumed to be generated
 instantaneously upon the event trigger, which can correspond to a
 simple flare, or one HXR 
pulse within a large complex flare. These eddies are characterised by 
a generation scale  $l_e$,
and speed $v_{e0}=b_0 v_{\rm A}$. The particle
 acceleration from the 
background plasma is determined by small-scale turbulence % intensity
characterised by the amplitude of the corresponding magnetic field
fluctuations $b B$. The growth of $b$ is driven by large-scale
eddies with growth rate $\tau_r^{-1} = v_{e0}/l_e=b_0/\tau_0$,
where $\tau_0=l_e/v_{\rm A}$ is the transit time of Alfv\'{e}n waves
through $l_e$. The turbulence starts to decay
once $b$ reaches $b_0$, and we adopt the Kraichnan
phenomenology with the decay time given by $\tau_d=l_ev_{\rm
A}/v_e^2=\tau_e v_{\rm A}/v_e$, where the eddy speed $v_e = b v_{\rm
A}$ and the eddy turnover time $\tau_e = l_e/v_e$ \citep{k65}. Then we have
\begin{equation} 
\dot{b}/b =\left\{
\begin{array}{ll}
\tau_r^{-1}= b_0/\tau_0\ \ \ \ &{\rm for\ the\ turbulence\ rise\ phase}\,, \\
-\tau_{d}^{-1}=-b^2/\tau_0\ \ \ \ &{\rm for\ the\ turbulence\ decay\
 phase}\,. \\ 
\end{array}
\right.
\label{dotd}
\end{equation}
This is the proposed basic equation for the turbulence
evolution. Considering the variety of flare triggers \citep{a02}, the
turbulence evolution can be much more complicated. These equations can be
modified wherever there are sufficient observational or theoretical
justifications. 

\begin{figure}[ht]
\vspace{-0mm}
\begin{center}
\includegraphics[width=7.0cm]{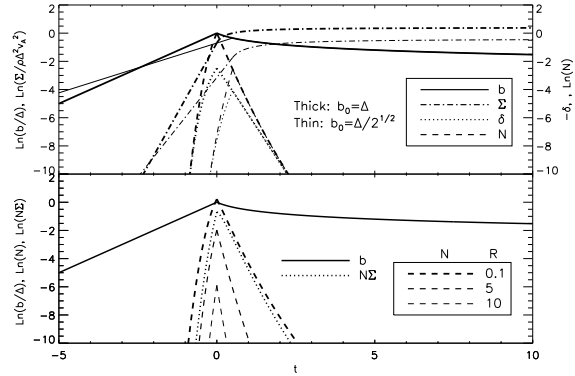}
\vspace{-0mm}
\caption{
 Evolution of elementary energy release events. The time unit is
 $\tau_0/\Delta$ and $\tau_0/\Delta^2$ for the rise and decay phase,
 respectively. Upper: the thick
 and thin lines are for $b_0=\Delta$ and
 $\Delta/2^{1/2}$, respectively.  The latter
 lines are shifted to the right by 
 $\tau_0/2\Delta^2$ so that the decay phase overlaps with
 the former. $N$ is for $E_0=10 k_{\rm B}T$ and normalised to the
 peak value for $b_0=\Delta$. 
%The weaker event also has a longer rise phase.
%The dot-dashed line shows the evolution of $\delta$, which reaches the
% minimum of $5/2$ for $b=\Delta$.
% For an electron thermal energy $k_{\rm B}T$ being equal
% to the transition energy between the thermal and nonthermal component
% $E_t$ for $b=\Delta$, the dashed line shows the
% evolution of electron density $N(E)$ at $E=10k_{\rm B}T$ with the
% normalisation being the corresponding value for $b=\Delta$.
%For the weak event with the
% model parameters the same as the strong one,
% the electron density and spectral index evolution are
% shown by the two lower thin solid lines for clarity. They are also
% shifted to the 
% right by $\tau_0/2\Delta^2$ so that the decay phase overlaps with the
% strong one. The three-dot-dashed lines show the correlation between
% $\delta$ (vertical axis) and $N(E)$. The logarithm of the normalised
% values of $N(E)$ are added by 20 and indicated by the horizontal
% axis. The upper and lower lines are for $E = 20$ and $10 k_{\rm B}T$,
% respectively.
Lower:  the dotted line is for
 $N\Sigma$. The solid line and
 normalisations of $N$ 
 and $\Sigma$ are the same as the upper panel. The dashed lines are the
 same as the thick-dashed line in 
 the upper panel except that
 from high to low the corresponding electron temperatures are 0.1, 5,
 and 10 times lower, respectively. 
%Note that the energy of
% nonthermal electron $E$ remains the same.
%The corresponding
% correlation between 
% $\delta$ and $N(E)$ are indicated by the three thin three-dot-dashed
% lines from low to high, respectively. The thick dashed and
% three-dot-dashed lines are the same as these in the left panel except
% that $N(E)$ is scaled by a factor of $\Sigma(t)/\rhob_0^2v_{\rm A}^2$
% to mimic the change of electron density through chromospheric
% evaporation.   
\label{lit}}
\end{center}
\end{figure}
We
consider the solution, where $b=b_0$ at $t=0$:
\begin{equation}
b(t) =b_0\left\{
\begin{array}{ll}
\exp(t/\tau_r)\ \ \ \ &{\rm for\ t\le0\ the\ turbulence\ rise\ phase}\,, \\
(1+2tb_0^2/\tau_0)^{-1/2}\ \ \ \ &{\rm for\ t>0\ the\ turbulence\
 decay\ phase}\,. \\ 
\end{array}
\right.
\label{delta}
\end{equation}
The solid lines in Figure \ref{lit} show the time
evolution of $b/\Delta$ for $b_0=\Delta$ 
%\footnote{The rise phase is relatively shorter than the decay phase for
%$\Delta<1$.}
(thick) and
$\Delta/2^{1/2}$ (thin), where $\Delta\le 1$ is an upper limit for
$b_0$, below which
the turbulence evolution is not significantly affected by nonthermal particles
(see discussions near the end of this section). 
%For the latter case, we also shift the peak time to
%$t=\tau_0/2\Delta^2$ so that the decay phase becomes identical to part
%of the decay phase of the former case.  
%With the above model, the
%turbulence decay is determined by the turbulence intensity and does not
%depend on the evolution history beforehand. 
%The decay phases are
%always identical later on. However, the rise phase is
%determined by the speed of large scale eddies produced by the event
%trigger. Weaker flares have longer rise phases.
The corresponding total amount of energy per unit volume dissipated
through small scale turbulence is given by
\begin{equation}
\Sigma(t) =2\rho v_{\rm A}^2\int^tb^2\tau_d^{-1}{\rm d t}^\prime=
\rho v_{\rm A}^2b_0^2\left\{
\begin{array}{ll}
(b_0/2)\exp(4t/\tau_r)\ \ \ \ &{\rm for\ t\le0}\,, \\
b_0/2+1-(1+2tb_0^2/\tau_0)^{-1}\ \ \ \ &{\rm for\ t>0}\,, \\
\end{array}
\right.
\label{energy}
\end{equation}
where $\rho$ is the mass density. It is indicated by the
dot-dashed lines in the Figure. The total energy dissipated through
turbulence is $(b_0/2+1)\rho v_A^2b_0^2$ with the first and
second term resulted from the rise and decay phase, respectively. Less
than $1/3$ of the total released energy is dissipated in the turbulence
rise phase. Most of the energy is dissipated in the much longer decay
phase. The rise of $\Sigma(t)$ approximately represents the rise of thermal
emission. % better than $b(t)$ does. 
%\footnote{In the calculations above, we assume that the flare starts at
%the infinite past with no small-scale turbulence initially. For more
%general cases, where the flare starts with some low level of small-scale
%turbulence, $b_i\ll b_0$, Equation (\ref{delta}) remains valid with the
%flare starting at $t_i = \ln(b_i/b_0)\tau_r$. The
%released energy given by equation (\ref{energy}) will be reduced by
%$\rho v_{\rm A}^2b_i^4/2b_0$, which is negligible.}

%\section{Stochastic Particle Acceleration}
%\label{acc}

%Once the turbulence reaches small scales, 
%comparable to the gyro-radii of the background particles, 
%Some of the background particles
%lose Coulomb coupling with others and be accelerated to very high
%energies. 
%It is possible that the large-scale turbulence may also be damped by
%some collective plasma effects. These processes likely energise
%particles in bulk resulting in plasma heating and may be considered for
%individual flare studies.  
We characterise the decoupling and acceleration of individual particles 
from the thermal background by small scale turbulence with
an acceleration timescale \citep{m97, p04}
\begin{equation}
\tau_{ac}=S_a \tau_t b^{-2}\,,
\end{equation}
%c^2/v_{\rm A}^2
where $S_a$ is a dimensionless coefficient describing the
acceleration by waves.
%, and
%$c$ is the speed of light and $\tau_t=l_e/v$, $v$ the particle speed. 
The energy loss time
$\tau_l(E)$ of high-energy particles through Coulomb collisions with a
low temperature background 
is proportional to $E^{3/2}/n$, where $E\gg k_{\rm B} T$ is the particle
kinetic energy,  
$n$ and $T$ are the background particle density and
temperature, and $k_{\rm B}$ is the Boltzmann constant.
The particle distribution is nonthermal at energies, where 
%the acceleration time is shorter than the Coulomb energy loss time 
$\tau_{ac}(E)<\tau_l(E)$.
%\footnote{We assume that Coulomb collisions  are the dominant particle
%energy loss process in solar flares.}  
The transition from the non-relativistic thermal
to nonthermal component occurs at
\begin{equation}
E_t = 2[\pi \ln\Lambda
e^4 n \tau_{ac}/m^{1/2}]^{2/3}\,,
\end{equation} 
where $e$, $m$ are the particle charge
and mass, respectively, and $\ln\Lambda\simeq 20$ for plasmas in the
solar corona, i.e., $\tau_{ac}(E_t) = \tau_l(E_t)$ \citep{p04}. 

The most natural way to produce a power-law high-energy particle
distribution with an adjustable spectral index is in terms of a
particle loss process as proposed 
originally by \citet{f49} for cosmic rays. For solar flares, this
particle loss is due to 
%decoupling of the particle from the background particles and 
escape from the energy release site through
spatial diffusion. We use
the scattering time 
\begin{equation}
\tau_{sc}=S_s \tau_tb^{-2}\,
\end{equation}
to characterise this
spatial diffusion, % of energetic particles,
% through the turbulent energy release site, 
where $S_s$ is the second coefficient describing the particle
scattering by waves. The corresponding particle escape 
time 
\begin{equation}
\tau_{esc} = l_0^2/v^2\tau_{sc}\,,
\end{equation}
where $l_0>l_e$ is the length of the
energy release site. The
electron acceleration timescale is much shorter than the flare duration.
%of the flare $\sim \tau_0/b_0^2$. 
One may consider the
steady-state solution.
% for the distribution of nonthermal particles. 
To
have a power-law distribution as commonly adopted to model the observed
nonthermal emissions, $\tau_{ac}$ and $\tau_{esc}$ need to have the same energy
dependence, i.e., $\tau_{ac}\tau_{sc}\propto 1/v^2$.
 $\tau_{sc}$ and $\tau_{ac}$ may have different energy dependence
\citep{p04}. For simplicity, we assume that their energy dependence is
the same.
For non-relativistic particles, as considered in the paper,
$v\propto E^{1/2}$, therefore
$\tau_{ac}\propto E^{-1/2}$, and both $S_a$ and
 $S_s$ are independent of
$v$. \footnote{Other assumptions of the energy dependence of $\tau_{ac}$
and $\tau_{sc}$ will lead to different energy dependence of
$\tau_{esc}$, which determines the flux of escaping particles and can be 
constrained by observations \citep{b06}. For example, with the standard
quasi-linear theory, $\tau_{ac}/\tau_{sc}\propto (v/v_{\rm A})^2$. It
can be shown that $\tau_{esc}$ is independent of $E$ and $S_a\propto 1/S_s
\propto v$.}

In the high-energy range,
where $\tau_{ac}\ll\tau_1$,
% the acceleration time is much shorter than the Coulomb
     % collisional energy loss time, 
the kinetic equation for nonthermal particle
distribution $f(p)$ is given by  
\begin{equation}
{\partial f\over \partial t} = {1\over p^2}{\partial\over \partial
 p}p^4
%(
\tau_{ac}^{-1}
%-\tau_c^{-1})
{\partial\over \partial
 p}f
%+{\partial\over \partial p}(p\tau_c^{-1}f) 
- {f\over \tau_{esc}}+Q\,,
\end{equation}
where $p$ is the particle momentum, and $Q$, a source term, exists at
low energies. In the
steady-state,
\begin{equation}
f(p)\propto p^{-2-(4+\tau_{ac}/\tau_{esc})^{1/2}}\,.
\end{equation}
The corresponding energy distribution
$
N(E) \propto E^{-1/2-(1+\tau_{ac}/4\tau_{esc})^{1/2}}
%(9/4+\tau_{ac}/\tau_{esc})^{1/2}-1/2
%\propto r_g^2l_0^{-2}{b}^{-4}c^2v_A^{-2}
\,,
$
and the flux of escaping particles
$
F(E) \propto N(E)/\tau_{esc} \propto E^{-(1+\tau_{ac}/4\tau_{esc})^{1/2}}\,.
$
%For the convergence of the energy flux, 
For the convergence of energy flux carried away by nonthermal particles,
the index of $F(E)$ must be
greater than 2, i.e., 
\begin{equation}
\tau_{ac}>12\tau_{esc}\,, \, \ \ {\rm}\  \ \ b^4<S_aS_s
%(c/v_{\rm A})^2
(l_e/l_0)^2/12\equiv \Delta^4\,\,.
\end{equation}
%The energy flux carried away by nonthermal particles from the
%acceleration site diverges 
For $b\ge \Delta$, %and in such a case,
one must
consider the relativistic effect and/or the effect of nonthermal particles
on the turbulence cascade 
and damping \citep{b09}. We consider the linear regime, where
nonthermal particles do not affect the turbulence cascade. $\Delta$ is
an upper limit for the linear model to be valid.
%It is interesting to note that the spectral index of electrons escaping from
%the flare site and observed near the Earth appear to have a low bound
%of 2.5, in agreement with this requirement \citep{kk07}.

\section{Application to Solar Flares}
\label{dis}

Electron acceleration during flares is better observationally
constrained than that 
for ions. 
%Since electrons and ions energy exchange through Coulomb collision
%has a much longer timescale than their Coulomb collision times with
%their kinds(a factor of $m_i/m_e$ longer than
%electron-electron collision and $(m_i/m_e)^{1/2}$ longer than ion-ion
%collision), we will also ignore this process.
We next apply the model to electron acceleration in flaring loops with ions
treated as a background of positive charges, which only changes through
large scale hydrodynamic processes. \footnote{This is an
assumption. Ion acceleration can be 
considered in appropriate theoretical and/or observational contexts.}
Then the length of the energy 
release site $l_0$ has a lower limit of the size of coronal looptop
sources \citep{x08} and an upper limit of the length of flaring loops
\citep{l06}. 
Although high-energy electrons must escape from the acceleration site to
produce a power-law distribution with the above SA model, charge
neutrality of the acceleration site requires the
total number of electrons remains the same as positive charges at
$N_0=l_0An$ (e.g., due to a return current), which gives the
normalisation for the electron distribution at the acceleration site $N(E)$.
 One of the
key purposes of the paper is to introduce a simple phenomenological model
to extract the energy release rate through turbulence from
observations. Instead of solving the full kinetic equation for
the electron distribution over the whole energy, i.e., from the Coulomb
collision dominated low-energies to collisionless high-energies
\citep{p04, g05}, we assume that the electron 
distribution is thermal below $E_t$, a power law above it, and
continuous at $E=E_t$. Then
$N(E)=N_0 g(E)$, where $g(E)$ is the normalised distribution function,
i.e., $\int g(E){\rm d} E=1$ and
\begin{equation}
g(E) = {g_0(E_t/k_{\rm B}T, \delta)\over (k_{\rm B}T)^{3/2}}\left\{
\begin{array}{ll} 
E^{1/2}\exp({-E/k_{\rm B} T}) & \textrm{for $E<E_t$}\,,\\
E_t^{1/2}\exp({-E_t/k_{\rm B}T})(E/E_t)^{-\delta} & \textrm{for
 $E\ge E_t$}\,, 
\end{array}  \right.
\end{equation}
where 
\begin{equation}
\delta = 1/2+(1+\tau_{ac}/4\tau_{esc})^{1/2}\,,
\end{equation}
\begin{equation}
g_0(r, \delta) =
 [r^{3/2}\exp(-r)/(\delta-1)-r^{1/2}\exp(-r)+\pi^{1/2}\mbox{erf}(r^{1/2})/2]^{-1
 }\,,    
\end{equation}
and $\mbox{erf}$ is the error function.
%We assume that 
%\begin{equation}
%\tau_{ac} = S_ab^{-2}
%c^2v_A^{-2}
%(l_e/v)\,,
%\end{equation} and 
%\begin{eqnarray}
%\tau_{sc}= S_s b^{-2}(l_e/v)\,,
%\end{eqnarray}
%where $S_a$, $S_s$ are two dimensionless energy independent coefficients
%characterise the interactions of electrons with the turbulence. Then
%\begin{eqnarray}
%\tau_{esc} &=& l_0^2b^2/S_s v l_e\,, \\
$E_t = (2\pi\ln\Lambda S_a l_e n)^{1/2}
%c
e^2/b 
%v_{\rm A}
\,,$
%\end{eqnarray}
%Clearly, $\delta$
%=S_aS_sc^4/\Omega^2b^4v_A^2l_0^2$ needs to be greater than $5/2$ if
%there is no high energy cutoff.\footnote{The escape of thermal electrons
%from the acceleration site can be treated as conduction. However,
%scattering by turbulence may reduce the diffusion of electrons near
%$E_t$, which may lead to a conductivity much lower than the Spitzer
%conductivity (Jiang et al. 2006).} Then we have
%$\tau_{ac}>12\tau_{esc}$ then gives  
%\begin{equation}
%S_aS_s>12b^4(v_{\rm A}/c)^{2}(l_0/l_e)^{2}\,.
%\end{equation} 
$\delta = 1/2 +(1+3\Delta^4/b^4)^{1/2}\,, 
r=E_t/k_{\rm B}T\equiv R(\Delta/b)\,$ with $R =(2\pi\ln\Lambda
S_a l_e n)^{1/2}e^2/k_{\rm B}T\Delta$.  
%The energy flux carried away by nonthermal electrons is given by 
%\begin{equation}
%\dot{\cal E}_{ne} = \int_{E_t}^\infty N(E)E\tau_{esc}^{-1}{\rm d}
%E={S_s  l_e n A v_t E_t \over b^{2} l_0}
%{n_0r^{3/2}\exp(-r)\over(\delta-5/2)}
%={N_0 v_t E_t \over \tau_{esc}v}
%{n_0r^{3/2}\exp(-r)\over(\delta-5/2)}
%\,.
%\end{equation}
%Then the electron
%acceleration efficiency 
%\begin{equation}
%\eta_{ne} = {\dot{\cal E}_{ne}\over \dot{\cal E}}=
%{S_s \tau_0 l_e n E_t v_t\over 2 \rho b^6 v_{\rm
%A}^2 l_0^2}{n_0r^{3/2}\exp({-r})\over(\delta-5/2)}\,. 
%\end{equation}

%The electron distribution at the energy release site 
$N$ is described
with four parameters: $N_0$, 
$T$, $E_t$, and $\delta$. For a given energy release event
discussed in Section \ref{turbulence} with $b_0$ given in units of
$\Delta$, the evolution of $\delta$ is 
determined as indicated by the dotted lines in Figure \ref{lit}. For
$b_0=\Delta$, $\delta$ reaches $5/2$ at the peak of $b$. $N_0$ and $T$
can be measured observationally and vary gradually as $\Sigma$ on a
timescale much longer than $\tau_{ac}$.
%, and the nonthermal particle density is less sensitive to them than to
%$\delta$. 
%depends on the turbulent heating, conductive and radiative cooling
%processes. 
%Detailed self-consistent treatment of its evolution is beyond the scope
%of this paper.  
We assume that they do not change during these
events.
For $E_t = k_{\rm B}T$ at 
$b=\Delta$ (the fiducial model), the evolution of $N$ at $E_0 = 10
k_{\rm B}T$ (normalised to the peak value) is 
indicated by the dashed line.
%, where it has been normalised to the peak value. 
Because $\delta$ has a strong dependence on $b$, the
nonthermal electron density at high energies is very sensitive to
$b$. A prominent peak of nonthermal electron numbers only appears near the
peak of $b$. The 
thin lines show another event with about 2
times lower released energy. As expected, the electron distribution
becomes softer than the previous model with a minimum of $\delta$ at
$1/2+13^{1/2}\simeq 4.1$. The peak 
density of nonthermal electrons is smaller by more
than one order of magnitude. The decrease of the peak density is more
prominent at even higher energies due to the softer spectrum. Therefore
strong and hard nonthermal 
emissions are expected only for strong events with $b_0$ close to
$\Delta$, and even with a relatively gradual evolution of the turbulence
intensity, an impulsive nonthermal component appears at the peak of the
energy dissipation. This may explain the bursty nature of the
nonthermal component as compared with the thermal component. The
dependence of $N(E_0)$ on $T$ is shown in the lower panel. There are
fewer nonthermal electrons for lower $T$. The
evolution of $T$ and $N_0$ will affect the 
quantitative details but not the impulsive nature of the nonthermal
density, which is mostly determined by $\delta(t)$.

\begin{figure}[ht]
\vspace{-0mm}
\begin{center}
\includegraphics[width=7.8cm]{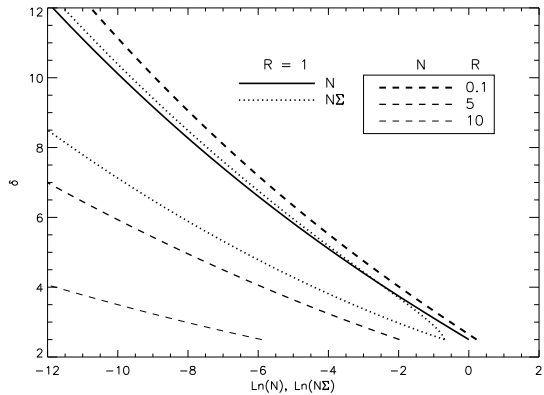} 
\vspace{-0mm}
\caption{
% Evolution of elementary energy release events and
Correlation between nonthermal electron spectral index and
 density. 
The solid line is for the fiducial model. 
%corresponds to the thick lines in the upper panel
% of Figure \ref{lit}. 
The other lines correspond to those in the lower panel of Figure
 \ref{lit}.   
\label{cor}}
\end{center}
\end{figure}

One of the most important observations of nonthermal emission is the spectral
SHS evolution of HXR pulses, which \citet{g04, g06} 
attributed to elementary electron acceleration events. The spectral
index and nonthermal electron density evolution of our model naturally
reproduces such a pattern. The thick solid line in Figure \ref{cor} shows
the correlation between $\delta$ and $N(E_0)$ of the
fiducial model.
% shown in the upper panel of Figure \ref{lit}. 
%In the left panel, $N$ is shown at $10$ and $20 k_{\rm B}T$. As expected, the
%latter changes more dramatically than the former. For the steady-state model, 
The correlation at a given energy is the same for the rise and
decay phase in agreement with observations of some HXR
pulses. %However, it is found that 
For most pulses, this correlation changes from
the rise to decay phase
% of most pulses has a higher nonthermal electron flux than the rise
% phase at a given $\delta$ 
\citep{g04}. 
This can be caused by chromospheric evaporation and/or plasma heating
due to the energy release $\Sigma$, which
changes the density and temperature of the background plasma. 
The dashed lines show how the correlation changes with $T$ and the dotted
line shows the correlation between $\delta$ and $N\Sigma$. 
%Due to continuous increase of $\Sigma$, $N\Sigma$ in the decay phase
%has a higher value than the rise phase at a given $\delta$. If the
%density increase of the acceleration region is proportional to the
%released energy $\Sigma$, this correlation can be compared with the
%observed correlation between nonthermal electron flux and spectral
%index during HXR pulses. Clearly t
The chromospheric evaporation processes must be modelled properly to
study this correlation.

\section{Conclusions}
\label{con}

Yohkoh and RHESSI observations of solar flares have revealed several
phenomena challenging the classical electron beam model: impulsive
soft X-ray 
emission from footpoint sources \citep{h94, h08}, continuous heating of the
coronal source after the impulsive HXR phase \citep{l97, j06}, and
more recently injection of electrons over a large coronal looptop
region \citep{x08}. All these point to the SA model, where electrons are
accelerated by magnetised turbulence in flaring loops \citep{p04}. The
acceleration site can be as compact as the observed coronal looptop
sources and may also extend over the whole loops if the turbulence is
transported along the loops quickly, e.g., in the form of plasma waves 
\citep{f08}. The latter may explain impulsive soft X-ray emission from
footpoints as electrons are injected at the footpoints
directly. In this paper, we propose a simple phenomenological SA 
model and show that the nonthermal electron density in the acceleration
site is very sensitive to the turbulence intensity. It therefore
provides a mechanism to produce impulsive emissions even with a
relatively gradual energy release process, which may account for the
continuous heating inferred from thermal source evolution after the
impulsive phase. With a simple model for the turbulence evolution, 
it also reproduces spectral SHS evolution of HXR pulses. Detailed
modelling of the plasma heating and chromospheric evaporation is needed
to quantify the flux and spectral index correlation.

Solar flares are multi-scale phenomena in terms of not only the size
and duration but also the amount of energy released by each flare. The
physical processes involved in the large-scale energy release process
can be scale dependent, which in combination with the variety of initial
and boundary conditions in the solar atmosphere is expected to lead to very
rich appearance \citep{a02}. However, the microscopic scale processes
of plasma 
heating and particle acceleration should be mostly determined by
properties of the background plasma in terms of temperature, density, and
large scale magnetic field. Besides Coulomb collisions among these
particles, particle interactions with the electromagnetic field
fluctuations during the energy release process may be parameterized with
an acceleration and scattering coefficient. We show in the paper that
these two coefficients, in combination with the turbulence intensity and
generation scale,
should be used to determine the electron  distribution in flaring
loops and applied to individual flares. The model may not only overcome
the %above mentioned 
observational challenges to the classical electron
beam model 
but also address the intriguing problem of the low-energy cutoff. The
two coefficients should not vary significantly. The turbulence intensity
and release scale $l_e$ are determined by the macroscopic processes of
flare triggers. 
%which leads to well defined particle acceleration efficiency for individual
%energy release event. 
Detailed analyses of flare X-ray emissions will
lead to quantitative constraints on the wave-particle interactions and
turbulence evolution and may also help to understand the
related large-scale processes.

\acknowledgements
This work is supported by the EU's SOLAIRE
Research and Training Network at the University of Glasgow
(MTRN-CT-2006-035484) and by Rolling Grant ST/F002637/1 from the UK's Science and Technology Facilities Council. We thank Hugh Hudson, Eduard Kontar, John
Brown, and Alec MacKinnon for helpful comments. 

{}


\begin{thebibliography}{}

%\bibitem[Alexander \& Daou (2007)]{a07}Alexander, D., \& Daou,
%		A. G. 2007, ApJ, 666, 1268

%\bibitem[Allred et al. (2005)]{a05} Allred, J. C., Hawley, S. L.,
%		Abbett, W. P., \& Carlsson, M. 2005, ApJ, 630, 573

\bibitem[Antonucci et al. (1984)]{a84}Antonucci, E., Gabriel, A. H.,
		\& Dennis, B. R. 1984, ApJ, 287, 917  

\bibitem[Aschwanden (2002)]{a02} Aschwanden, M. J. 2002, Space Science
		Reviews, 101, 1

\bibitem[Battaglia \& Benz (2006)]{b06} Battaglia, M., \& Benz,
		A. O. 2006, A\&A, 456, 751

\bibitem[Benz (1977)]{b77}Benz, A. O. 1977, ApJ, 211, 270

%\bibitem[Brown (1971)]{b71} Brown, J. C. 1971, Solar Phys. 18, 489 

\bibitem[Brown (1973)]{b73} Brown, J. C. 1973, Solar Phys. 31, 143 

%\bibitem[Brown \& Melrose (1977)]{b77} Brown, J. C., \& Melrose,
%		D. B. 1977, Solar Phys. 52, 117

\bibitem[Bykov \& Fleishman (2009)]{b09} Bykov, A. M., \& Fleishman,
		G. D. 2009, ApJ, 692, L45 

\bibitem[Dennis \& Zarro (1993)]{d93}Dennis, B. R., \& Zarro,
		D. M. 1993, Solar. Phys., 146, 177

\bibitem[Fermi (1949)]{f49}Fermi, E. 1949, Phys. Rev. 75, 1169

\bibitem[Fletcher \& Hudson (2008)]{f08} Fletcher, L., \& Hudson,
		H. S. 2008, ApJ, 675, 1645

\bibitem[Galloway et al. (2005)]{g05} Galloway, R. K., MacKinnon, A. L.,
		Kontar, E. P., \& Helander, P. 2005, A\&A, 438, 1107

%\bibitem[Goldreich \& Sridhar (1995)]{g95} Goldreich, P., \& Sridhar,
%		S. 1995, ApJ, 438, 763

\bibitem[Grigis \& Benz (2004)]{g04}Grigis, P. C., \& Benz, A. O. 2004,
		A\&A, 426, 1093

%\bibitem[Grigis \& Benz (2005)]{g05}Grigis, P. C., \& Benz, A. O. 2005,
%		A\&A, 434, 1173

\bibitem[Grigis \& Benz (2006)]{g06}Grigis, P. C., \& Benz, A. O. 2006,
		A\&A, 458, 641

\bibitem[Hamilton \& Petrosian (1992)]{h92}Hamilton, R. J., \& Petrosian,
		V. 1992, ApJ, 398, 350

\bibitem[Hannah et al. (2008)]{h08} Hannah, I. G., Krucker, S., Hudson,
		H. S., Christe, S., \& Lin, P. R. 2008, A\&A, 481, L45

\bibitem[Holman et al. (2003)]{h03} Holman, G. D., Sui, L. H., Schwartz,
		R. A., \& Emslie, A. G. 2003, ApJ, 595, L97

\bibitem[Hudson et al. (1994)]{h94} Hudson, H., S., et al. 1994, ApJ,
		422, L25 

\bibitem[Jiang et al. (2006)]{j06}Jiang, Y. W., Liu, S., Liu, W., \& Petrosian,
		V. 2006, ApJ, 638, 1140


%\bibitem[Jiang et al. (2009)]{j09}Jiang, Y. W., Liu, S., \& Petrosian,
%		V. 2009, ApJ, in press.


%\bibitem[Kasparova et al. (2007)]{k07} Kasparova, J., Kontar, E. P., \&
%		Brown, J. C. 2007, ApJ, 466, 705

%\bibitem[Kontar \& Brown (2006)]{k06} Kontar, E. P., \& Brown,
%		J. C. 2006, ApJ, 653, L149

%\bibitem[Kontar \& Reid (2009)]{k09} Kontar, E. P., \& Reid,
%		H. A. S. 2009, ApJ, 695, L140

\bibitem[Kraichnan (1965)]{k65} Kraichnan, R. H. 1965, Phys. of Fluids,
		8, 1385

%\bibitem[Krucker et al. (2007)]{kk07} Krucker, S., Kontar, E. P.,
%		Christe, S., \& Lin, P. R. 2007, ApJ, 663, L109

%\bibitem[Krucker et al. (2008)]{k08} Krucker, S., et al. 2008, A\&A
%		Rev. 16, 155

\bibitem[Li et al. (1997)]{l97}Li, P., McTiernan, J. M., \& Emslie,
		A. G. 1997, ApJ, 491, 395

\bibitem[Liu et al. (2006)]{l06} Liu, W., Liu, S., Jiang, Y. W., \&
		Petrosian, V. 2006, ApJ, 649, 1124

%\bibitem[Miller et al. (1996)]{m96} Miller, J. A., LaRosa, T. N., \&
%Moore, R. L.,  1996, \apj, 461, 445

%\bibitem[Miller \& Roberts (1995)]{m95} Miller, J. A., \& Roberts, D. A.
%1995, \apj, 452, 912

\bibitem[Miller et al. (1997)]{m97} Miller, J. A., et al. 1997,
		J. Geophys. Res. 102, 14631

%\bibitem[Montgomery \& Tuner (1981)]{m81} Montgomery, D., \& Turner,
%		L. 1981, Phys. Fluids, 24, 825

\bibitem[Neupert (1968)]{n68} Neupert, W. M. 1968, ApJ, 153, L59

%\bibitem[Paesold et~al. (2003)]{p03} Paesold, G., Kallenbach, R., \&
		%Benz, A. O. 2003, \apj, 582, 495 

%\bibitem [Petrosian \& Donaghy (1999)]{p99} {Petrosian, V., \& Donaghy,
		%T. Q.\ 1999, \apj, 527, 945} 

\bibitem [Petrosian \& Liu (2004)]{p04} Petrosian, V., \& Liu, S.\
	 2004, \apj,  10, 550

%\bibitem[Pryadko \& Petrosian (1997)]{p97}Pryadko, J. M., \& Petrosian,
%V. 1997, ApJ, 482,  74 

\bibitem[Saint-Hilaire \& Benz (2005)]{s05} Saint-Hilaire, P., \& Benz,
		A. O. 2005, A\&A, 435, 743

%\bibitem[Sui et al. (2007)]{s07} Sui, L. H., Holman, G. D., \& Dennis,
%		B. R. 2007, ApJ, 670, 862

\bibitem[Veronig et al. (2005)]{v05} Veronig, A. M., et al. 2005, ApJ, 621, 482

 
\bibitem[Xu et al. (2008)]{x08}	Xu, Y., Emslie, A. G., \& Hurford,
		G. J. 2008, ApJ, 673, 576	

%\bibitem[Zharkova \& Gordovskyy (2006)]{z06} Zharkova, V. V., \&
%		Gordovsky, M. 2006, ApJ, 651, 553

\end{thebibliography}
\end{document}